\documentclass[twocolumn,letter]{jpsj2} 

\def\runauthor{H. Fukazawa {\it et al.}}

\title{%
Possible Multiple Gap Superconductivity with Line Nodes\\
in Heavily Hole-Doped Superconductor KFe$_{2}$As$_{2}$\\ 
Studied by $^{75}$As-NQR and Specific Heat
}

\author{%
Hideto~Fukazawa$^{1,4}$\thanks{E-mail address: hideto@nmr.s.chiba-u.ac.jp}, 
Yuji~Yamada$^{1}$,
Kenji~Kondo$^{1}$,
Taku~Saito$^{1}$,
Yoh~Kohori$^{1,4}$, 
Kentarou~Kuga$^{2}$, 
Yosuke~Matsumoto$^{2}$, 
Satoru~Nakatsuji$^{2}$, 
Hijiri~Kito$^{3,4}$, 
Parasharam~M.~Shirage$^{3}$,
Kunihiro~Kihou$^{3,4}$,
Nao~Takeshita$^{3,4}$,
Chul-Ho Lee$^{3,4}$, 
Akira~Iyo$^{3,4}$, 
and Hiroshi~Eisaki$^{3,4}$
}

\inst{%
$^{1}$Department of Physics, Chiba University, Chiba 263-8522, Japan\\
$^{2}$Institute for Solid State Physics (ISSP), University of Tokyo, Kashiwa 277-8581, Japan\\
$^{3}$National Institute of Advanced Industrial Science and Technology, Tsukuba 305-8562, Japan\\
$^{4}$JST, TRIP, Chiyoda-ku, Tokyo 102-0075, Japan\\
}

\recdate{\today}

\abst{%
We report the $^{75}$As nuclear quadrupole resonance (NQR) and specific heat measurements of 
the heavily hole-doped superconductor KFe$_{2}$As$_{2}$ 
(superconducting transition temperature $T_{\rm c} \simeq 3.5~$K). 
The spin-lattice relaxation rate $1/T_{1}$ in the superconducting state exhibits 
quite gradual temperature dependence with no coherence peak below $T_{\rm c}$. 
The quasi-particle specific heat $C_{\rm QP}/T$ shows small specific heat jump which is about 30\% of 
electronic specific heat coefficient just below $T_{\rm c}$. 
In addition, it suggests the existence of low-energy quasi-particle excitation 
at the lowest measurement temperature $T=0.4$~K$\simeq T_{\rm c}/10$. 
These temperature dependence of $1/T_{1}$ and $C_{\rm QP}/T$ can be explained by multiple nodal superconducting gap scenario 
rather than multiple fully-gapped $s_{\pm}$-wave one within simple gap analysis. 
}

\kword{%
KFe$_{2}$As$_{2}$, multiple superconducting gap, nuclear quadrupole resonance, specific heat
}

\begin{document}
\maketitle



After the discovery of superconductivity in F-doped LaFeAsO with a superconducting transition temperature $T_{\rm c}=26$~K~\cite{Kam1}, 
K-doped (hole-doped) BaFe$_{2}$As$_{2}$ was reported as the first oxygen-free iron-pnictide superconductor with $T_{\rm c} = 38~$K~\cite{Rot2}. 
The crystal structure of Ba$_{1-x}$K$_{x}$Fe$_{2}$As$_{2}$ is of the ThCr$_{2}$Si$_{2}$-type. 
We performed $^{75}$As-NMR measurements of Ba$_{1-x}$K$_{x}$Fe$_{2}$As$_{2}$ ($T_{\rm c} \simeq 38~$K)~\cite{Fuk6}, 
which also exhibits the first-order antiferromagnetic (AF) ordering associated with a structural phase transition. 
Our results clearly revealed that the coexistence of AF and superconducting (SC) states~\cite{Rot3,HCh1} is not a microscopic one but a phase separation, 
which is related with the fact that the structural transition is intrinsically of the first order~\cite{Hua1}.
Hence, the suppression of the structural phase transition rather than carrier doping 
seems to have a key role in achieving superconductivity in so-called 122 systems. 
Recent reports of pressure-induced superconductivity of BaFe$_{2}$As$_{2}$ ($T_{\rm c} \simeq 13~$K)~\cite{Take2} 
and SrFe$_{2}$As$_{2}$ ($T_{\rm c} \simeq 34~$K)~\cite{Kot3,Matb1} under hydrostatic pressure are consistent with this consideration. 

The important feature of 
Ba$_{1-x}$K$_{x}$Fe$_{2}$As$_{2}$ is that 
superconductivity occurs even for $x=1$~\cite{Rot3,HCh1,Sas1}, 
though $T_{\rm c}$ itself is much lower ($T_{\rm c} \sim 3.5~$K) than the optimum $T_{\rm c}$. 
This also implies the less essentialness of carrier doping for the occurrence of superconductivity 
in the 122 systems compared to high-$T_{\rm c}$ cuprates. 
However, looking at the phase diagram of Ba$_{1-x}$K$_{x}$Fe$_{2}$As$_{2}$ in ref.~\ref{Rott}, 
$T_{\rm c}$ once decreases to zero from maximum $T_{\rm c}$ side toward $x = 0.75$ with increasing $x$, 
but remains lower value ($T_{\rm c} \sim 10$~K) above about $x = 0.75$. 
Nearly the same tendency of the phase diagram is also seen in Sr$_{1-x}$K$_{x}$Fe$_{2}$As$_{2}$~\cite{Sas1}. 
This is probably related with the disappearance or the shrinkage of the electron-like Fermi surface 
around the M point in the Brillouin zone by hole doping, 
which is indeed observed by angle resolved photoemission spectroscopy (ARPES)~\cite{Sat1} 
or confirmed by the band calculation in KFe$_{2}$As$_{2}$~\cite{Sin1}. 
Furthermore, this might bring the change of the SC symmetry. 
In iron arsenide superconductors, it has been proposed that the multiple fully-gapped $s_{\pm}$-wave Cooper paring is 
preferable from many experimental and theoretical approaches~\cite{Maz1,Kur1,Kur2,Ike1,Nom1,Shib1,Naka1,Yas1}. 
On the other hand, the nodal-line SC symmetry scenario is proposed for superconductivity in LaFePO 
due to the modification of the Fermi surface~\cite{Kur2,Fle1,Yam1}. 
In this sense, KFe$_{2}$As$_{2}$ is a suitable candidate 
for verifying the general tendency of SC symmetry in Fe-based superconductors, 
since it is potentially cleaner than other substituted SC compounds and SC even with extremely modified Fermi surface.
In this Letter, we report the $^{75}$As nuclear quadrupole resonance (NQR) 
and specific heat measurements of KFe$_{2}$As$_{2}$ ($T_{\rm c} \simeq 3.5~$K). 
Characteristic temperature $T$ dependence of spin-lattice relaxation rate $1/T_{1}$ and quasi-particle specific heat $C_{\rm QP}/T$ 
can be understood by multiple nodal SC gap scenario.



Polycrystalline KFe$_{2}$As$_{2}$ was synthesized by a high-temperature and high-pressure method~\cite{Fuk6}. 
X-ray diffraction analysis revealed that the sample was of nearly single phase. 
The obtained lattice parameters, $a=3.846~{\rm \AA}, c=13.87~{\rm \AA}$, were consistent with the previous reports~\cite{Rot3,HCh1}. 

 \begin{figure}
  \centering
  \includegraphics[width=8cm]{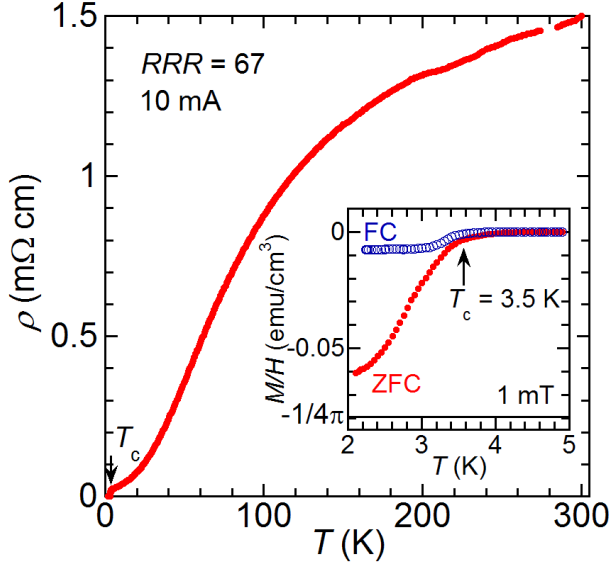}
  \caption{
  (Color online) Resistivity of KFe$_{2}$As$_{2}$. 
  The inset shows the magnetic susceptibility of KFe$_{2}$As$_{2}$ in the paramagnetic state. 
  Solid and open symbols denote the data obtained after zero-field-cooling (ZFC) and field-cooling (FC), respectively. 
  }
 \end{figure}

In order to check the sample quality, we measured resistivity $\rho$($T$) by a standard four-probe method (Fig.~1). 
The residual resistivity ratio ($RRR$) is $\rho$(300~K)/$\rho$(4.2~K) = 67, 
which is comparable with the reported value (= 87) for single crystal of KFe$_{2}$As$_{2}$ ($T_{\rm c}$ = 2.8~K)~\cite{Ter1}. 
The onset of $T_{\rm c}$ is about 4.0~K and the $\rho$ becomes nearly zero at 3.5~K. 
We also determined the $T_{\rm c}$ and SC volume fraction with a commercial SQUID magnetometer (the inset of Fig.~1). 
To reduce demagnetization effect, we used the plate-shape sample with 3.0$\times$3.0$\times$1.0~mm$^{3}$ 
and applied magnetic field parallel to the largest plane. 
The $T_{\rm c}$ of the sample is about 3.5~K and the SC volume fractions is 80\%. 
These results indicate that the quality of our sample is reasonably good enough and 
that the superconductivity in KFe$_{2}$As$_{2}$ is bulk property of the sample. 

The NQR experiment on the $^{75}$As nucleus ($I=3/2$, $\gamma/2\pi = 7.292$~MHz/T) was carried out 
using phase-coherent pulsed NQR spectrometers. 
The samples were crushed into powder for use in the experiments. 
The measurement above 1.4~K was performed using a $^{4}$He cryostat, 
and between 0.3 and 1.2~K with a $^{3}$He refrigerator. 
We measured specific heat by a thermal relaxation method between 0.4 and 10~K 
with a commercial calorimeter.


 \begin{figure}
  \centering
  \includegraphics[width=8cm]{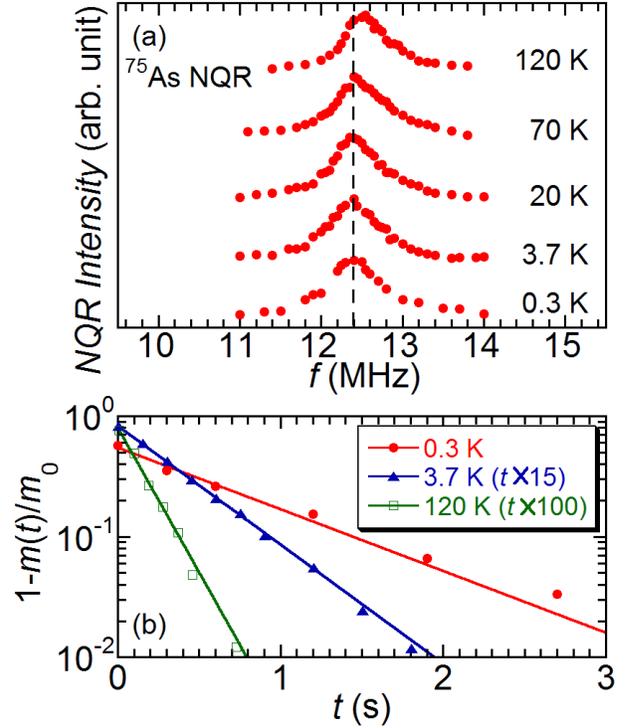}
  \caption{
  (Color online) (a) $^{75}$As NQR spectra of KFe$_{2}$As$_{2}$ at various temperatures.
  (b) Nuclear magnetization recovery curves of KFe$_{2}$As$_{2}$ at 0.3, 3.7 and 120~K. 
  Solid lines denote the fitting curve using the formula written in the text.
  }
 \end{figure}

In Fig.~2(a), we show the $^{75}$As-NQR spectra of KFe$_{2}$As$_{2}$ at various $T$'s. 
Clear single peak signal was successfully observed. 
With decreasing $T$, the spectral center decreases and remains nearly constant below about 70~K. 
Below this temperature, the NQR frequency $\nu_{Q}$
was 12.4~MHz. 
The principal axis of the electric field gradient is along the crystal $c$-axis 
since the As site has a local four-fold symmetry around the $c$-axis. 
With increasing $x$, the $\nu_{Q}$ in Ba$_{1-x}$K$_{x}$Fe$_{2}$As$_{2}$ increases from $\nu_{Q} = 2.2$~MHz ($x=0$)~\cite{Fuk2,Ktg1}
to 12.4~MHz ($x=1$) through intermediate value $\sim$5~MHz ($x=0.4$)~\cite{Fuk6}. 
The full width at half maximum ($FWHM$) remains nearly constant above $T_{\rm c}$ and is about 740$\pm$20 kHz. 
This is about half magnitude of linewidth of oxygen-deficient LaFeAsO$_{1-\delta}$~\cite{Muk1} and F-doped LaFeAsO$_{1-x}$F$_{x}$~\cite{Kawa1}. 
The $FWHM$ slightly increases below $T_{\rm c}$ and becomes about 850~kHz at 0.38~K. 
However, we cannot conclude that this is due to magnetic order 
since the increase of the width is too small. 

In fig.~2(b), we show nuclear magnetization recovery curves of KFe$_{2}$As$_{2}$ at 0.3, 3.7 and 120~K. 
All the obtained recovery curves followed single exponential curve expected for $^{75}$As NQR ($I = 3/2$)~\cite{Mac1}, 
$$1-\frac{m(t)}{m_{0}} = \exp\left( -\frac{3t}{T_{1}}\right),$$
where $m(t)$ and $m_{0}$ are nuclear magnetizations after a time $t$ from the NQR saturation pulse 
and thermal equilibrium magnetization. 
Small deviation of data points from the fitting curve for 0.3~K may be due to the distribution of $T_{\rm c}$. 
However, its fitting error is within marker size in Fig.~3. 
We checked the absence of sample heat up by changing the NQR pulse width and power. 

 \begin{figure}
  \centering
  \includegraphics[width=8cm]{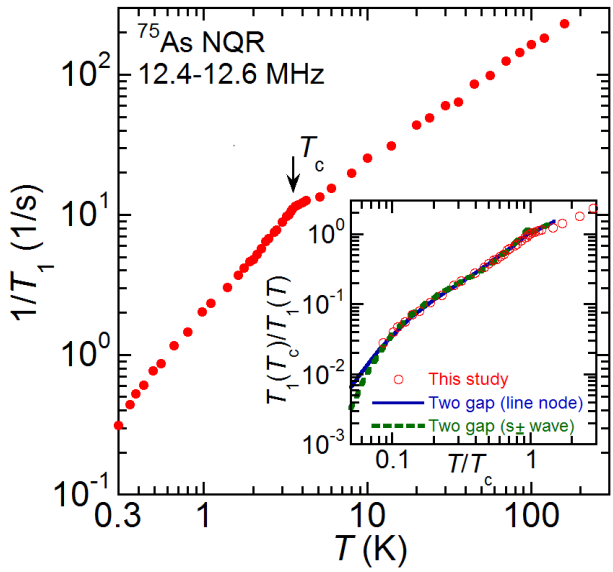}
  \caption{
  (Color online) $1/T_{1}$ of KFe$_{2}$As$_{2}$. 
  The inset shows the normalized results of the experiment and analysis. 
  We assumed two independent SC gap model for nodal gap or full gap. 
  }
 \end{figure}

In Fig.~3, we show the $T$ dependence of $1/T_{1}$ of KFe$_{2}$As$_{2}$. 
In the normal state, $1/T_{1}$ follows $T^{0.8}$ between 5 and 160~K. 
This is slightly gradual compared to $T$ linear dependence of usual Korringa law, 
which suggests that antiferromagnetic fluctuation is much suppressed and 
that the system is nearly Pauli paramagnetic. 
The most striking feature of $1/T_{1}$ is $T$ dependence below $T_{\rm c}=3.5$~K. 
No coherence peak was observed just below $T_{\rm c}$, 
and the $1/T_{1}$ follows only $T^{1.4}$ between 0.6 K and $T_{\rm c}$. 
This is quite different from the $1/T_{1}$ below $T_{\rm c}$ in Ba$_{1-x}$K$_{x}$Fe$_{2}$As$_{2}$ 
with $T_{\rm c} \simeq 38~$K in which $1/T_{1}$ follows $T^{3}$-$T^{5}$ without coherence peak~\cite{Fuk6,Mata2,Yas1}. 
The $1/T_{1}$ of KFe$_{2}$As$_{2}$ more steeply decreases below about 0.6~K. 
Such kind of gradual $T$ dependence of $1/T_{1}$ in the SC state was observed in 
La$_{0.87}$Ca$_{0.13}$FePO~\cite{Nak2}.

 \begin{figure}
  \centering
  \includegraphics[width=6cm]{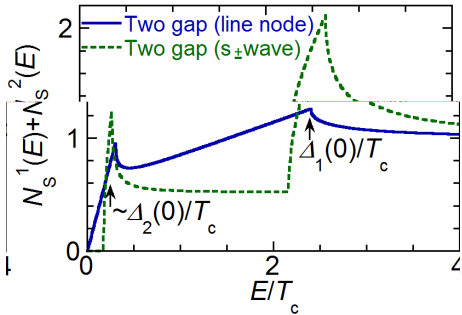}
  \caption{
  (Color online) Total density of states 
  used for the analysis. 
  }
 \end{figure}

In order to understand the nature of this anomalous $T$ dependence,
we analyzed the data by assuming simple two-independent SC-gap model with gap with line node or full gap ($s_{\pm}$ wave type). 
Similar analysis procedure was adopted in refs.~\ref{Yash},~\ref{Matano}. 
Larger two hole Fermi surfaces ($\alpha$ and $\beta$ in ref.~\ref{Sato}) are observed around the $\Gamma$ point by ARPES in KFe$_{2}$As$_{2}$~\cite{Sat1}. 
These Fermi surfaces have different sizes of gap in  Ba$_{1-x}$K$_{x}$Fe$_{2}$As$_{2}$ ($T_{\rm c} \simeq 38~$K)~\cite{Shib1,Naka1,Yas1,Mata2}, 
and are consider to have main contribution to total density of states (DOS) at the Fermi level in KFe$_{2}$As$_{2}$~\cite{Sat1,Sin1}. 
Therefore, it is natural to analyze two SC gap model. 
Furthermore, we assumed two SC gap symmetries because these symmetries are experimentally and theoretically 
favorable for iron-based superconductors~\cite{Maz1,Kur1,Kur2,Ike1,Nom1,Shib1,Naka1,Yas1,Fle1,Mata2}. 
The $1/T_{1}$ in the SC state is proportional to, 
$$\frac{1}{T_{\rm 1}} \propto \sum_{i=1,2}n_{i}^{2}\int_{0}^{\infty}\{ N_{\rm S}^{i}(E)^{2}+M_{\rm S}^{i}(E)^{2}\} f(E)\{ 1-f(E)\} dE,$$
where $N_{\rm S}^{i}(E)$, $M_{\rm S}^{i}(E)$, $f(E)$ are the DOS, the anomalous DOS 
arising from the coherence effect of Cooper pairs, and the Fermi distribution function, respectively. 
$n_{i}$ represents the fraction of DOS of the $i$-th gap and $n_{1}+n_{2}=1$. 
For both cases of SC gaps, we assumed that the integral of $M_{\rm S}^{i}(E)$ becomes zero because of the sign-changing SC gaps. 
For the full gap case, we averaged $N_{\rm S}^{i}(E)$ around $E \sim \Delta_{i}(T)$ 
with the width $2\delta_{i} (<\Delta_{i}(T))$ in order to reduce the coherence peak~\cite{Heb1,Shi1}. 
Here, $\Delta_{i}(T)$ is the SC gap; for simplicity, ($\theta, \varphi$) dependence is given 
by $\Delta_{i}(T,\theta,\varphi)=\Delta_{i}(T)\cos\theta$ for nodal gap and $\Delta_{i}(T,\theta,\varphi)=\Delta_{i}(T)$ for full gap. 
Concrete function for the averaged $N_{\rm S}^{i}(E)$ is described in ref.~\ref{Hebel}.
This procedure corresponds to consider the anisotropy of the SC gaps or finite lifetime of Cooper pairs arising from pair breaking~\cite{Amb1}.
In order to distinguish the nodal gap model and the full gap model, 
we utilized the above procedure different from that described in refs.~\ref{Yash},~\ref{Naga}, 
in which finite DOS appears at the Fermi level in the full gap model. 
The clear difference of total DOS is depicted in Fig.~4 by using the parameters described below. 

 \begin{table}[t]
  \caption{Analyzed SC gaps, fraction of DOS, and width of broadening of gap for the experimental $1/T_{1}$.}
  \label{t1}
  \begin{center}
   \begin{tabular}{@{\hspace{\tabcolsep}\extracolsep{\fill}}ccccc} \hline
   Type of gap & $2\Delta_{1}(0)/T_{\rm c}$  & $2\Delta_{2}(0)/T_{\rm c}$ & $n_{1}$ & $\delta_{1}/\Delta_{1}$ \\ \hline
   Line node & 4.8  & 0.60 & 0.45 & -\\
   Fully gapped $s_{\pm}$ & 4.8  & 0.44 & 0.48 & 0.1 \\ \hline
   \end{tabular}
  \end{center}
 \end{table}

In the inset of Fig.~3, we show the normalized results of the experiment and analysis. 
In Table.~I, we summarized the obtained parameters of $2\Delta_{i}(0)/T_{\rm c}$, $n_{1}$, and $\delta_{1}/\Delta_{1}$ (for the full gap model). 
The gap parameters are nearly the same for each model and the fitting to the experimental results is good enough. 
This suggests that the $^{75}$As-NQR down to the lowest measurement temperature $\sim T_{\rm c}/10$ 
cannot solely determine the SC symmetry of KFe$_{2}$As$_{2}$. 
The larger gap is of moderate strong coupling while the smaller gap is of quite weak coupling. 
This indicates that the smaller gap is induced by the emergence of the larger gap. 
Note that the single gap analysis ($n_{1}=1$) did not entirely work for both SC symmetries. 

 \begin{figure}
  \centering
  \includegraphics[width=8cm]{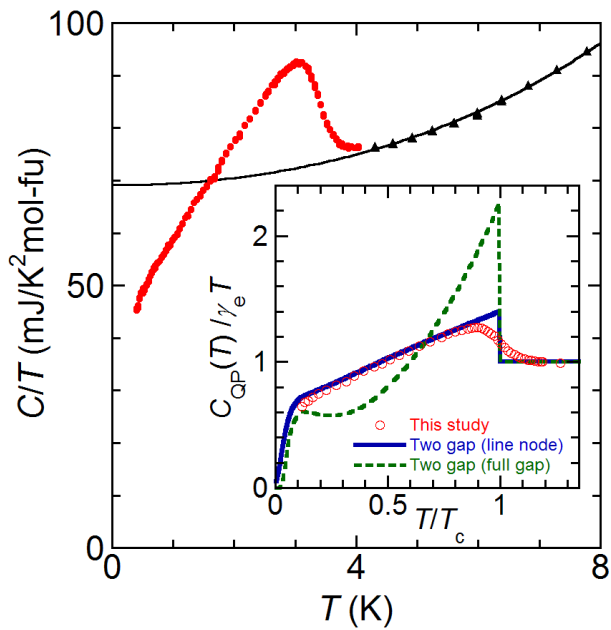}
  \caption{
  (Color online) Specific heat divided by temperature $C(T)/T$ of KFe$_{2}$As$_{2}$. 
  Solid curve represents the fitting ($C_{{\rm fit}}/T$) for the data between 4 and 10~K (the data below 8~K is shown). 
  The fitting function is described in text. 
  The inset shows the normalized results of experiment and numerical calculation. 
  We assumed two independent SC gap model for nodal gap or full gap and used the parameters summarized in Table.~I. 
  }
 \end{figure}

In Fig.~5, we show specific heat divided by temperature $C(T)/T$ of KFe$_{2}$As$_{2}$. 
Clear specific heat jump was observed, 
which again indicates the bulk nature of superconductivity. 
The midpoint of the jump is 3.4~K. 
In order to estimate the electronic-specific-heat coefficient $\gamma_{\rm e}$ and the lattice contribution, 
we performed the fitting ($C_{{\rm fit}}/T$ $ = \gamma_{\rm e}+\beta T^{2}+\varepsilon T^{4}$) for the data between 4 and 10~K (a solid curve in Fig.~5. 
The obtained $\gamma_{\rm e}$ is 69.1(2) mJ/K$^{2}$mol, 
which is comparable with 63.3 mJ/K$^{2}$mol for single crystal of Ba$_{0.6}$K$_{0.4}$Fe$_{2}$As$_{2}$~\cite{Mu1}. 
Note that the Schottky specific heat of $^{75}$As is negligible since the calculated Schottky specific heat 
from the NQR frequency $\nu_{Q}=12.4$~MHz becomes significant below about 0.3~K. 

By subtracting the lattice contribution $\beta T^{2}+\varepsilon T^{4}$, we obtained the quasi-particle specific heat $C_{\rm QP}/T$. 
The most obvious feature of $C_{\rm QP}/T$ is that the specific heat jump is only about 30\% of $\gamma_{\rm e}$. 
One of reasons is the broadening of the SC transition arising from the small distribution of $T_{\rm c}$. 
However, by taking into account this broadening, the specific heat jump is roughly expected to be at most 60\% of $\gamma_{\rm e}$. 
Another important feature is the finite $C_{\rm QP}/T \simeq 45$~mJ/K$^{2}$mol-fu even at the lowest measurement temperature $\sim T_{\rm c}/10$. 
This suggests the existence of low-energy quasi-particle excitation. 

In the inset of Fig.~5, we compared the normalized results of experiment and numerical calculation. 
Here, we again assumed simple two-independent SC-gap model with gap with line node or full gap. 
Quasi-particle specific heat is given as follows, 
$$\frac{C_{\rm QP}}{T} \propto \sum_{i=1,2}\frac{n_{i}}{T^{3}}\int_{0}^{\infty}N_{\rm S}^{i}(E)\left\{ E^2-\frac{T}{2}\frac{d\Delta_{i}(T)^{2}}{dT}\right\}$$
$\qquad\qquad\qquad\qquad\qquad\qquad\qquad \times f(E)\{ 1-f(E)\} dE.$\\
To calculate the specific heat, we used the parameters summarized in Table.~I. 
Clearly, the calculated result of the nodal SC gap model is 
in good agreement with the experimental result except for the shape around the SC transition. 
Especially this calculated result has quite small specific heat jump and exhibits the low-energy quasi particle excitation. 
Therefore, we may conclude that the multiple gap superconductivity with gap with line node 
is realized in KFe$_{2}$As$_{2}$ within the simple two gap analysis. 
The $E$ linear part between 0.5 and 2 in the total DOS for nodal gap model in Fig.~4 
gives rise to wide $T$ linear part in $C_{\rm QP}/T$ between $T_{\rm c}/5$ and $4T_{\rm c}/5$. 
Note that such correspondence between $C_{\rm QP}/T$ and total DOS can be 
seen in the results of specific heat (Fig.~7 in ref.~\ref{Mu})
and $1/T_{1}$ (Fig.~4(d) in ref.~\ref{Yash}) of Ba$_{0.6}$K$_{0.4}$Fe$_{2}$As$_{2}$. 
It is also suggestive that the specific heat jumps divided by $\gamma_{\rm e}$, 
$\Delta C_{\rm QP}/\gamma_{\rm e}T_{\rm c}$, in other typical multi gap superconductors 
are small ($\Delta C_{\rm QP}/\gamma_{\rm e}T_{\rm c} \simeq 0.75$) for nodal-gap Sr$_{2}$RuO$_{4}$~\cite{Nis1} 
and reasonably large ($\Delta C_{\rm QP}/\gamma_{\rm e}T_{\rm c} \simeq 1.2$) for full-gap MgB$_{2}$~\cite{Bou1}. 
In our present analysis, the rapid decrease in $1/T_{1}$ and $C_{\rm QP}/T$ is expected below $T_{\rm c}/10$ 
below which the influence of residual DOS becomes significant. 
Further studies below this temperature using single crystals are needed. 

From the present study, there is tendency that iron-pnictide superconductors with intrinsically lower $T_{\rm c} \sim 10$~K 
including FeP-based compounds~\cite{Fle1,Yam1} have nodal SC gap. 
This is supported by theory from the view point of modified Fermi surfaces and the change of favorable wave vector for superconductivity~\cite{Kur2}, 
though its band calculation is based on rather lightly-doped LaFeAsO and LaFePO. 
Moreover, it is suggested that there is SC symmetry change at $x \sim 0.75$ $(=3/4)$ 
in the phase diagram of $A_{1-x}$K$_{x}$Fe$_{2}$As$_{2}$ ($A$ = Ba, Sr). 
Hence, it is required from both experimental and theoretical aspects to reveal 
which Fermi surface or wave vector is responsible for superconductivity in KFe$_{2}$As$_{2}$ and 
clarify the properties of the possible phase boundary at $x \sim$ 3/4.


In summary, we performed the $^{75}$As NQR and specific heat measurements of 
the heavily hole-doped superconductor KFe$_{2}$As$_{2}$ ($T_{\rm c} \simeq 3.5~$K). 
The $1/T_{1}$ in the normal state reflects nearly Korringa-like $T$ dependence. 
The $1/T_{1}$ in the SC state exhibits quite gradual $T$ dependence with no coherence peak. 
The quasi-particle specific heat $C_{\rm QP}/T$ shows small specific heat jump just below $T_{\rm c}$. 
In addition, it suggests the existence of low-energy quasi-particle excitation at $T=0.4$~K$\simeq T_{\rm c}/10$. 
These $T$ dependence of $1/T_{1}$ and $C_{\rm QP}/T$ can be explained by multiple nodal superconducting gap scenario 
rather than multiple fully-gapped $s_{\pm}$-wave one which is the most plausible scenario 
to describe SC state in Ba$_{1-x}$K$_{x}$Fe$_{2}$As$_{2}$ with $T_{\rm c}\simeq 38$~K. 
Further studies including theory and other experimental methods (ARPES, de Haas-van Alphen effect, penetration depth etc) 
especially using single crystal at lower temperatures are required to more precisely determine the SC gap symmetry.  
Moreover, Knight shift measurement below $T_{\rm c}$ is also urgently required to determine the spin part symmetry of KFe$_{2}$As$_{2}$.


The authors thank K. Kuroki, H. Ikeda, T. Nomura, K. Deguchi for valuable discussion and suggestion. 
This work is supported by a Grant-in-Aid for Scientific Research on KAKENHI from the MEXT and the JSPS, 
Innovative Areas "Heavy Electrons" (No. 20102005) from the MEXT, 
Global COE program of Chiba University.

\end{document}